\documentclass[sigconf]{acmart}
\AtBeginDocument{%
  }

\usepackage{balance}
\setcopyright{acmlicensed}
\copyrightyear{2018}
\acmYear{2018}
\acmDOI{XXXXXXX.XXXXXXX}
\acmConference[Conference acronym 'XX]{Make sure to enter the correct
  conference title from your rights confirmation email}{June 03--05,
  2018}{Woodstock, NY}
\acmISBN{978-1-4503-XXXX-X/2018/06}

\begin{document}

\title{Making Databases Searchable with Deep Context}

\author{Alekh Jindal, Shi Qiao, Shivani Tripathi, Niloy Debnath, Kunal Singh, Pushpanjali Nema, Sharath Prakash, Aditya Halder, Ronith PR, Sadiq Mohammed, Abdul Hameed, Karan Hanswadkar, Ayush Kshitij, Sarthak Bhatt, Rony Chatterjee, Jyoti Pandey, Christina Pavlopoulou, Ravi Shetye\vspace{0.2cm}}
\email{research@tursio.ai}
\affiliation{%
  \vspace{0.2cm}
  \institution{Tursio}
  \city{Bellevue}
  \country{USA}
  \vspace{0.2cm}
}

% \affiliation{%

% \affiliation{%

% \affiliation{%

% \affiliation{%

% \affiliation{%

% \affiliation{%

% \affiliation{%

\renewcommand{\shortauthors}{Trovato et al.}

\begin{abstract}

  Databases are the most critical assets for enterprises, and yet they remain largely inaccessible to people who make the most important decisions. In this paper, we describe the Tursio search platform that builds an abstraction layer, aka semantic knowledge graph, over the underlying databases to make them searchable in natural language. Tursio infuses large language models (LLMs) into every part of the query processing stack, including data modeling, query compilation, query planning, and result reasoning. This allows Tursio to process natural language queries systematically using techniques from traditional query planning and rewriting, rather than black-box memorization. We describe the architecture of Tursio in detail and present a comprehensive evaluation on production workloads, and synthetic and realistic benchmarks. Our results show that Tursio achieves high accuracy while being efficient and scalable, making databases truly searchable for non-expert users.

\end{abstract}

\received{20 February 2007}
\received[revised]{12 March 2009}
\received[accepted]{5 June 2009}

\maketitle

\section{Introduction}

Databases are the sources of truth for enterprises. 
Yet, retrieving information from databases requires users to write SQL queries, which the vast majority of non-experts struggle with.
Generative AI promises to make it a level playing field by empowering the non-expert users in natural language, i.e., they can focus on ``what'' they want and not ``how'' to get it. SQL was already supposed to do that, but over the years it has become too complex to be used by non-experts. Large language models (LLMs) can now understand those non-experts directly in natural language and generate SQL queries (text-to-SQL). Unfortunately, general purpose LLMs are unaware of internal data within enterprises and prone to errors that could be disastrous~\cite{sun-times-ai-misinformation, mcdonalds-ai-drive-thru-white-castle, bbc-ai-mistakes}.

% Therefore, it is very hard to make them work in practice and 

% Chicago Sun-Times and Philadelphia Inquirer published summer reading recommendations of books that don't exist~\cite{sun-times-ai-misinformation}.
% Or, the McDonald's AI-powered drive-thru ordering system that kept adding more Chicken McNuggets to their order, reaching eventually to 260~\cite{mcdonalds-ai-drive-thru-white-castle}. 
% As ludicrous as these examples are, recent BBC report highlights an even more surprising trend: people are now getting paid to fix mistakes made by AI, with some claiming that up to $90\%$ of their work involves correcting these errors~\cite{bbc-ai-mistakes}.

% Accuracy is arguably the number one challenge in generative ai; 

% particularly,unacceptable for enterprises: how do they know when something is not correct
% often boils down to ai being only as good as the data it operates on;
% need to feed AI with facts and often the facts are stored in structured databases

Prompting better context is one way of improving the accuracy of general purpose LLMs.
For example, Databricks Genie allows users to create knowledge stores~\cite{databricks-genie-knowledge-store}, provide sample SQLs~\cite{databricks-genie-sql-queries-functions}, and provide other instructions that could be used for prompting. 
Likewise, Snowflake Cortex allows users to configure semantic views~\cite{snowflake-views-semantic-overview}, create verified query repositories~\cite{snowflake-cortex-verified-query-repository}, and provide custom instructions~\cite{snowflake-cortex-custom-instructions}.
However, users are responsible for managing table and column descriptions, synonyms, join relationships, among other instructions to improve accuracy.
They must also choose the sample questions, e.g., up to 100 questions in Databricks, along with their SQLs that are unambiguous and helpful for improving the accuracy. This involves a significant amount of effort that keeps accumulating over time to cope with new data, new models, and non-determinism.

Fine-tuning is the other alternative to improve SQL generation accuracy.
For example, Google Gemini recently topped the single-model BIRD benchmark by curating gold-standard datasets and running supervised fine-tuning to produce a stronger SQL specialist, combined with self-consistency inference~\cite{google-gemini-database-understanding}.
While such general-purpose fine-tuning improves the model's SQL fluency broadly, it does not encode the enterprise-specific context---proprietary schemas, business logic, domain terminology---that is essential for accurate query generation over internal databases.
Lamini takes a different approach by {\it memorizing} facts from specific databases for more targeted retrieval.
However, enterprise data keeps changing, and memorized facts need to be refreshed continually.
More fundamentally, memorizing facts or query patterns does not compose: the number of possible query plans grows combinatorially with schema size, making exhaustive memorization impractical.

% - fine-tuning is the holy grail for gen AI on enterprise data; better for determinism
% - however, it still suffers from accuracy
% - primarily because it tends to miss critical facts
% - better prompts and context to shift the weights
% - memory tuning to make loss near zero
% - in fact, lamini has text-to-SQL as a use case
% - they can achieve 95\% accuracy but recommend a small set of queries (20-30) to memorize
% - essentally, memorize large number of facts

% - memorization is a problem since facts are often derived, i.e., they dont have 1-1 mapping to answers
% - databases transform facts using operators that are stitched into query plans
% - memorizing facts is then akin to caching all possible query plans, which is very large
% - this is no different for text-to-SQL: the model needs to memorize all combinations of SQL transforms

\begin{figure}[!t]
  \includegraphics[width=0.475\textwidth]{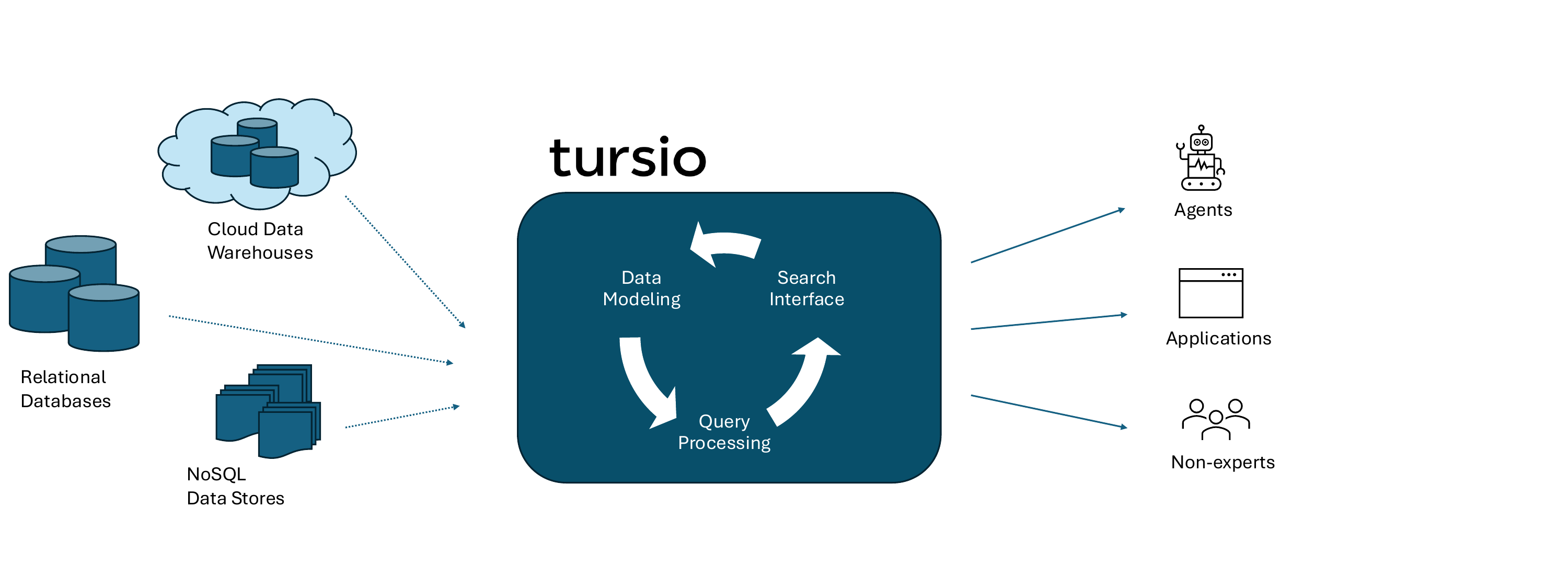}
  \caption{Tursio platform to make databases searchable for non-experts, applications, and agents.}  
  \Description{Tursio Overview}
  \label{fig:tursio}
\end{figure}

\begin{figure*}[!t]
  \includegraphics[width=\textwidth]{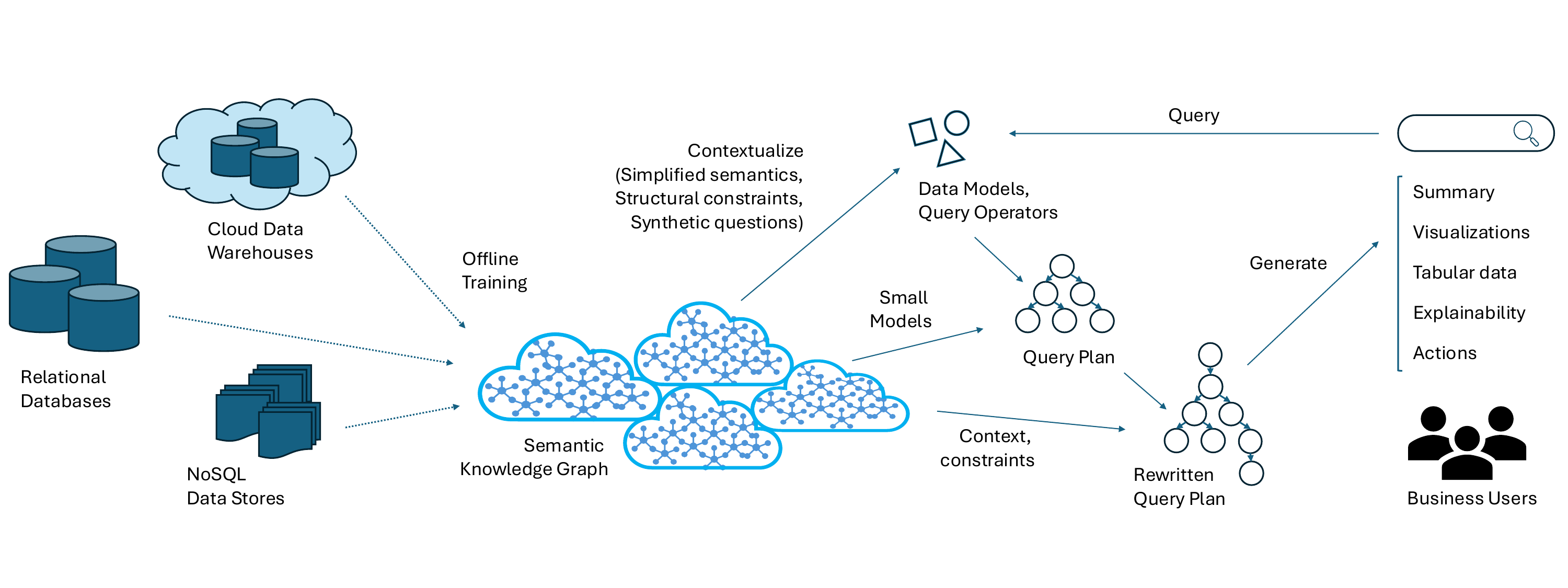}
  \vspace{-0.6cm}
  \caption{Detailed architecture of the Tursio search platform.}
  \Description{Tursio detailed architecture}
  \label{fig:tursio-overview}
  \vspace{-0.2cm}
\end{figure*}

Fundamentally, both prompting and fine-tuning ``black box" natural language queries into guesswork rather than processing them systematically.
This explodes the effort needed to achieve the required levels of accuracy and applicability. Furthermore, non-expert users care about the final results with SQL generation being the internal step that they cannot verify anyways.  
In this paper, we present Tursio, a brand new search platform to process natural language queries over databases and generate responses directly in natural language, as illustrated in Figure~\ref{fig:tursio}. The key idea is to connect to all databases, infer a semantic knowledge graph, contextualize user intent into the right data models using the knowledge graph, and process the queries systematically using techniques from traditional query planning and rewriting.
Along the way, Tursio infuses LLMs into every part of the query processing stack, including data modeling, query compilation, query planning, and result reasoning. Philosophically, this mirrors the idea from our prior work on learned query optimization~\cite{jindal2021microlearner, learnedCardinality, steeringOptimizers}: rather than relying on a single monolithic model, decomposing the problem into fine-grained steps---each augmented with a targeted model---yields better accuracy and debuggability.

In summary, our contributions are as follows:

\begin{enumerate}
  \item We present the Tursio search platform, a modern query engine to process natural language queries over databases and generate responses directly in natural language. (Section~\ref{sec:overview})
  \item We describe a novel LLM-powered approach to infer a semantic knowledge graph from the underlying databases, including things like semantics, relationships, constraints, and others. (Section~\ref{sec:semantics})
  \item We describe our technique to contextualize user intent using the semantic knowledge graph, and infer the relevant data models and query operators. (Section~\ref{sec:intent})
  \item We describe a systematic approach to generating query plans for the natural language queries, using techniques from traditional query planning and rewriting. (Section~\ref{sec:planning})
  % \item We describe our prompt-based query rewriting techniques to improve the expressivity of the query plans to match them as closely as possible to the user intent. (Section~\ref{sec:rewriting})
  \item We outline the Tursio search interface that provides a rich set of features for users to interact with the results, including explainability, visualizations, sharing, and others. (Section~\ref{sec:interface})
  \item We discuss the enterprise readiness of Tursio from the perspectives of security, privacy, compliance, governance, and others. (Section~\ref{sec:readiness})
  \item We show a detailed evaluation of Tursio on production workloads, test workloads from our SaaS platform, and synthetic and realistic benchmarks. (Section~\ref{sec:evaluation})
  \item Finally, we discuss the road ahead for Tursio and generative AI for databases in general. (Section~\ref{sec:roadahead})
\end{enumerate}

% - gen AI model: question -> answer
% - thats not how database query engines work: they leverage metadata to compile queries into valid query plans, before optimizing and running it in the best possible way
% - need reasoning based query processing and black box memorization based
% - black boxing explodes the training set needed for the required levels of accuracy, sufficient examples (question, answer) pairs, i.e., the query plans, reducing the applicability to very narrow data sets and use cases
% - throwing data to the problem 
% - low accuracy, inconsistency (change change every time), performance (hard for model to reason optimizations and depends on the kind of queries, optimized or unoptimized, in the training dataset)

% - our previous work addressed a similar problem in query optimization: 
% - giant monolithiic models fail to learn complex query optimizers 
% - instead, we build large number of small models, called micromodels, to learn every step in the process
% - apply a similar approach for building a generative query engine: 
% - question -> tokenize -> parse -> metadata -> operator -> operands -> operator tree -> query plan -> optimized query plan -> answer

\section{Tursio Overview}
\label{sec:overview}

Tursio provides non-expert business users a modern way of interacting with databases in natural language.
Rather than generating SQL queries that most users cannot verify anyways, Tursio exposes a simplified search interface that abstracts all complexity behind the scenes. Figure~\ref{fig:tursio-overview} shows the detailed architecture of the Tursio search platform, with three main components: inferring the semantic knowledge graph, generating query plans from natural language queries, and providing the search interface to business users.

First of all, Tursio connects to all major databases (including SQL Server, Azure SQL, PostgreSQL, MySQL, and Oracle), data warehouses (including Fabric, Snowflake, Databricks, BigQuery, and Teradata), and even NoSQL data stores (including Cassandra, CosmosDB, Excel, and Power BI Vertipaq). We leverage SQLAlchemy-based connectors to tap into existing databases and make them searchable in a single place without moving any data around. Tursio pushes all queries down to the respective databases and surfaces responses in the same search interface. Such a design decouples search from the database platform, allowing it to scale independently with newer data sources over time.

Once connected, Tursio infers a semantic knowledge graph from the underlying databases. The knowledge graph captures the semantics of the data, including things like name simplification, join inference, dimensions and measures, table and column descriptions, ontologies, data and value types, value aspects, valid aggregations, default measures, aliases, personally identifiable information (PII), sample values, and others. This knowledge graph automates the traditional data modeling step that is often manual, tedious, and error-prone. Instead, Tursio infers the knowledge graph automatically, using LLMs, and uses it to identify the right data models for user queries. It also keeps updating the knowledge graph over time as the underlying databases evolve.

When a user issues a natural language query, Tursio first contextualizes the user intent using the semantic knowledge graph. This involves identifying the right data models that match the user intent using techniques like hash prediction, join graph traversal, semantic lookup, and LLM-based judgment. Tursio also identifies the operations needed to answer the query using pre-generated sample questions and semantic matching from the selected data models. Thereafter, Tursio generates a query plan by parsing query fragments into meaningful operators, grounding the parsed operators into valid ones using metadata, generating a tree of well-formed operators, and further rewriting the query plan to be more expressive in answering the user questions as closely as possible.

Finally, Tursio provides a rich search interface for business users to interact with the results. The interface supports features like auto-prompting, result reasoning (including agentic reasoning for larger datasets), explainability (including interpreted and non-interpreted results), hints for similar operations and questions, tabular interactions, visualizations, data models, sharing (including PDF and link), saving, history, applying external context, and so on. The interface abstracts all complexity behind the scenes, allowing non-expert users to focus on their business questions without worrying about SQLs or query plans.

In the rest of the paper, we describe each of the above components in detail before presenting a detailed evaluation.

\section{Semantic Knowledge Graph}
\label{sec:semantics}

\begin{figure}[!t]
  \includegraphics[width=0.475\textwidth]{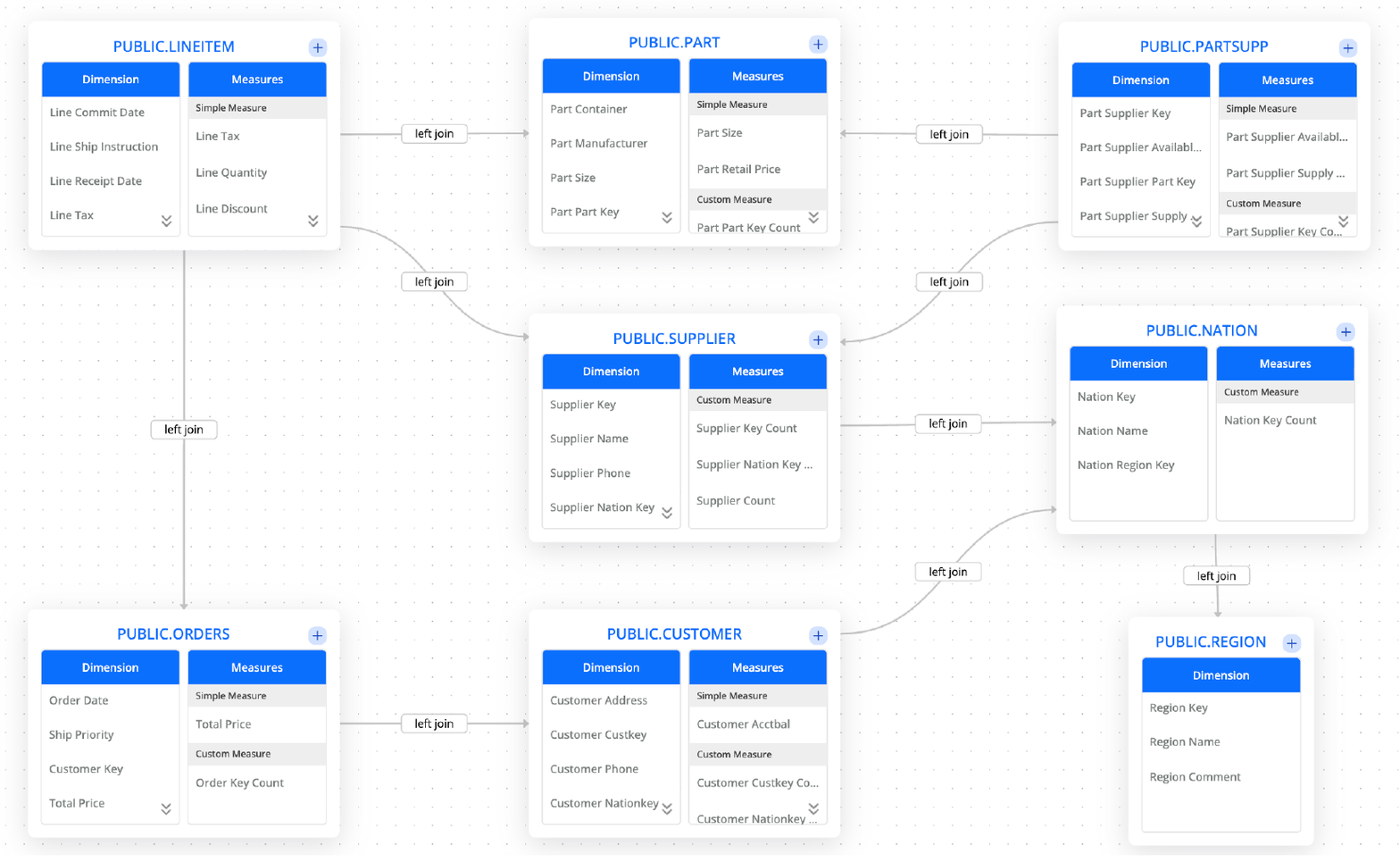}
  \caption{Semantic knowledge graph for TPC-H database.}
  \Description{Semantic Model}
  \label{fig:semantic-model}
\end{figure}

Traditionally, data or semantic modeling is the process of mapping databases to the business users. Unfortunately, this requires manual effort, based on the business requirements, to model only the relevant portions of the database. Moreover, the BI analysis keep adding more data with more business requirements over time. Tursio inverts this process by inferring a semantic knowledge graph over {\it all} data and then identifying the relevant portions to use as data models at query time. Figure~\ref{fig:semantic-model} shows an example semantic knowledge graph inferred by Tursio over the TPC-H database.

\vspace{0.2cm}
{\bf Profiling.}
Once a database is connected to Tursio, users can select the tables to be included in the semantic knowledge graph. We first profile the columns in each of these tables. We collect fixed-sized samples for each column, ranging from 100k to 1M rows, depending on the table sizes.
Column values in production databases often have data quality issues, and so we apply a data cleaning step to gather more representative ones. Specifically, we remove outliers by filtering out the values that lie outside the inter-quartile range (IQR) for numeric values and whose length lie outside the inter-quartile range for string values. This prunes extreme low or high values, or values that are likely to be typos or invalid.
We note that profiling focuses on representative value distributions and may not capture all data quality patterns, e.g., when the same information exists in both raw and cleaned columns. However, the column descriptions and LLM-based reasoning in subsequent steps help disambiguate such cases by comparing column semantics beyond their value distributions alone.
We also collect statistics, e.g., counts, count distincts (full or approximate based on table sizes), min/max, average, stddev, nulls, etc.

\vspace{0.2cm}
{\bf Nomenclature.}
Once profiled, we classify columns into dimension and measure columns using schemas, statistics, and LLMs, e.g., ``customer name'' is a dimension while ``order amount'' is a measure. Furthermore, columns names are often abbreviated and hard to understand, e.g., ``c\_acctbal'' instead of ``customer account balance''. Tursio uses LLMs to simplify such names into more human-friendly ones that could be used as context.
We further generate descriptions for columns and tables and columns using LLMs, based on their names, statistics, and sample values. These descriptions help disambiguate the user intent and generate more relevant responses. 
Tursio also identifies personally identifiable information (PII) in columns using LLMs and excludes them during query planning to avoid accidental exposure of sensitive data. 
Finally, Tursio generates compact aliases for tables and columns names that are too verbose for better grounding.

\vspace{0.2cm}
{\bf Ontologies.} 
In addition to schema-level nomenclature, Tursio infers ontologies for columns using LLMs. It differentiates between four types of data, namely string, numeric, date, and boolean, and classifies their value types as description, identifier, or continuous.
Furthermore, it identifies hierarchical ontologies such as ICD, CPT, and LOINC codes, time series columns, columns representing metric names and values, identifiers which are not countable, valid aggregate functions for different measure columns (e.g., count can be summed while average cannot be aggregated), and so on. This helps generate query expressions that better represent the semantics of the data. Finally, we also infer value aspects such as units, costs (e.g., dollar, pound, euro), percentages, and precision for numeric columns for response generation purposes. 

\vspace{0.2cm}
{\bf Join Inference.}
Tursio identifies key-foreign key relationships between tables using an inclusion dependency discovery algorithm~\cite{Rostin2009AML}.
We first detect the primary keys for each table using a combination of statistics (approximate count distinct over successive samples) and LLM reasoning (resolve among comparable candidates). Then, we identify candidates for inclusion dependencies by assigning a weighted score of schema similarity (name and data type), value inclusion (approximate count distinct overlap over successive samples), sample sizes, and so on. Finally, we validate these candidates using LLM as a judge. Thus, we reduce the space deterministically using heuristic-based pruning before leverage LLMs over high quality candidates for the final decision. This also helps scale join inference to larger schemas with many tables.

\vspace{0.2cm}
{\bf Custom Measures.}
Tursio supports custom measures, i.e., users can provide custom SQL expressions that will be inferred when compiling natural language queries. 
It also creates COUNT(DISTINCT) custom measures for all identifier columns, i.e., ensure that identifies are not duplicated when counting them.
Tursio further supports importing custom measures from existing BI tools like Power BI by uploading their VPAX files containing the DAX expressions.

\vspace{0.2cm}
Overall, the semantic knowledge graph captures a holistic understanding of the database automatically, giving users a good starting point for searching their data. Users can still extend or modify the knowledge graph anytime, but they are not trapped in manually creating it in the first place.

% \section{Context Generation}
\section{Contextualizing User Intent}
\label{sec:intent}

Tursio processes natural language queries by first contextualizing the user intent to relevant portions of the semantic knowledge graph. This involves two main steps: (i)~identifying the input tables (data models) needed to answer the query, and (ii)~identifying the operations involved to answer the query.

\vspace{0.2cm}
{\bf Identifying Inputs.}
Tursio trains a hash-based predictor to deterministically map keywords in the user query to relevant tables in the semantic knowledge graph. This helps prune the search space significantly before applying other techniques. We leverage both schema and data, and apply standard NLP transformations like removing stopwords, stemming, lemmatization, and others for reliable hash prediction. In case of no match, we fall back to semantic search, via a vector index, on the schema.
Once we identify the initial set of inputs, we then traverse the join graph to identify all valid paths containing those inputs. Depending on the user question, we further expand the initial set of inputs along their valid join paths using LLMs. We pick the longest yet the most relevant join path, and use LLM as a judge to resolve any ambiguities.
By combining deterministic hash-based predictor with systematic join graph traversal and LLM adjudication, Tursio identifies the right parts of the knowledge graph to focus on.

  %  (hash predictor, join graph traversal, semantic lookup, LLM judge)
  
\vspace{0.2cm}
{\bf Identifying Operations.}
In addition to identifying the input tables, Tursio also identifies the operations needed to answer the user query. We pre-generate a large number of sample questions over various tables, covering a wide range of operations, including projection, filtering, aggregations, group by, order by, joins, and so on. These synthetically generated questions are semantically correct and provide a grounding for the LLMs, without requiring users to construct them manually. At query time, we first narrow down the sample questions to those containing the inputs identified above. Then, we apply semantic matching between the user query and the filtered sample questions and provide their corresponding operations (projection, filtering, aggregations, group by, order by, etc.) as context for the LLM to fragment the user query into operations. Thus, we extract the user intent based on the closest interpretation to what exists in the knowledge graph. 

To illustrate, a filter query {\it ``Show demographics for males after 2012''} can have the following context generated automatically:

\begin{small}
\begin{verbatim}

Question: List the demographics details for males after 2009.
Filters: [Gender='Male', Registration_Date>2009]

Question: List demographics details for Spaniards between Jan 05, 
2018 and Aug 28, 2009.
Filters: [Ethnicity='Spaniard', Registration_Date>=2018-01-05,
 Registration_Date<=2009-08-28]

Question: List demographics details for SSN not in 157549937 
and 155485548 between 2018 and 2009
Filters: [SSN!=157549937, SSN!=155485548, 
Registration_Date>=2018, Registration_Date<=2009]

\end{verbatim}
\end{small}

\section{Natural Language Query Planning}
\label{sec:planning}

\begin{figure}[!t]
  \includegraphics[width=0.475\textwidth]{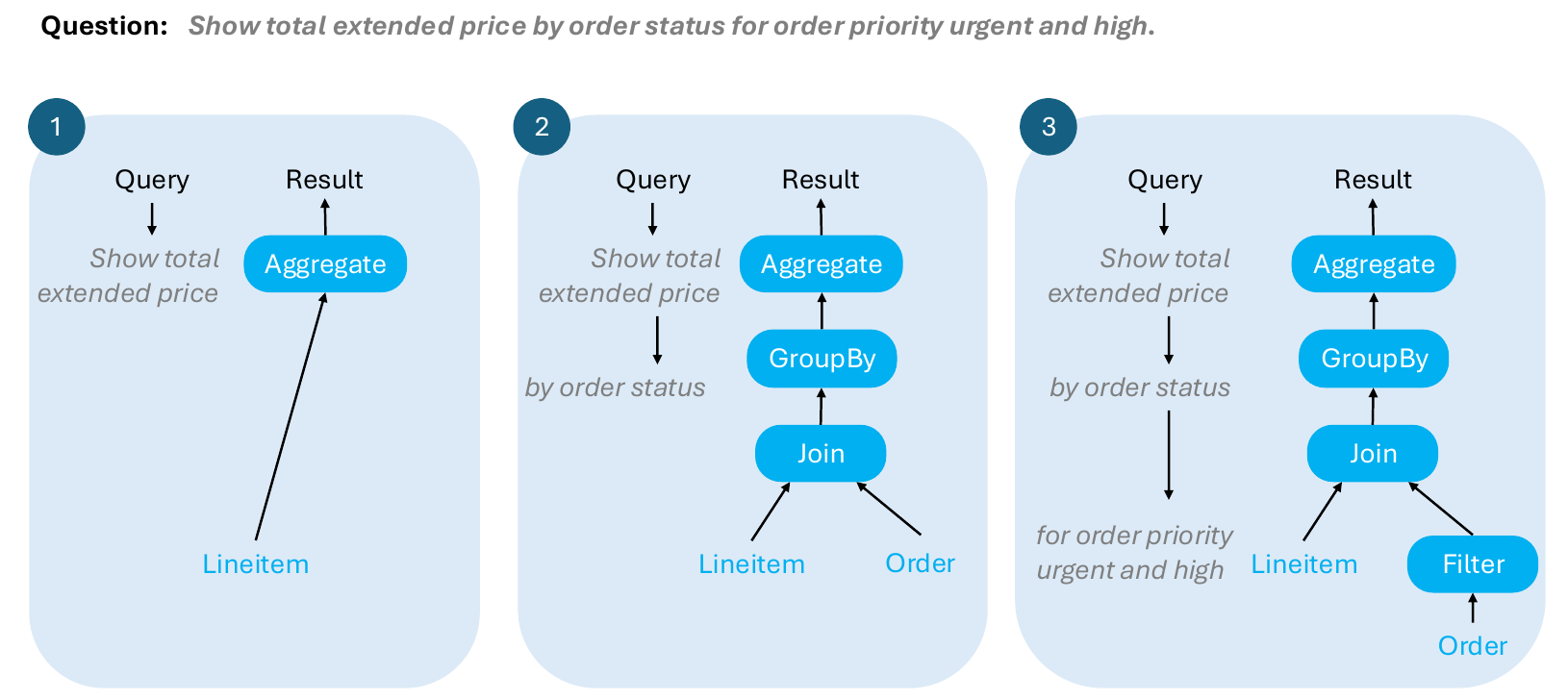}
  \caption{Natural language query planning in multiple steps compiling: (1) aggregate extended price in lineitem, (2) grouping by order status and implicit join with orders, and (3) filtering for order priority high and urgent.}
  \Description{Natural Language Query Planning}
  \label{fig:query-planning}
\end{figure}

Tursio generates query plans from natural language queries using a systematic approach, as illustrated in Figure~\ref{fig:query-planning}. The key idea is to break down the query planning into several steps, infusing LLMs into every step to handle the complexity of natural language.

A natural question is why not simply prompt an LLM with the relevant schema context and let it generate SQL directly. While end-to-end approaches work for simpler queries, they suffer from compounding errors on complex enterprise schemas: the model must simultaneously resolve table selection, join paths, column mapping, filter extraction, and aggregation semantics in a single generation step, with no opportunity to catch intermediate mistakes. Tursio's multi-step approach constrains the problem at each stage: parsed operators are validated by deterministic ANTLR parsers, operands are grounded against the knowledge graph metadata, and the operator tree enforces relational algebra rules. Each step narrows the space of valid outputs for the next, making errors detectable and correctable rather than silently propagated into the final SQL. This decomposition also enables the query breakdown feature (Section~\ref{sec:interface}), where users can see exactly which parts of their question were interpreted and how.

The main steps are as follows:

% Tokenization
% - fragments that could refer to structured operations \\
% - few-shot learning: provide domain-specfic context of questions -> fragments\\
% - context templates with transfer learning to ground into specifc database instances\\
% - dimensions, measures, filters, group by, order by, limit, etc.\\

\vspace{0.2cm}
{\bf Parsing.}
Given the operator fragments from the previous step, Tursio uses robust parsers to parse them into meaningful operators. We leverage ANTLR-based parsers with generalized logic to handle the noise from LLMs. This ensures that the parsed operators are well-formed and can be further processed in the subsequent steps. For example, filters are parsed into a list of conditions (equality, inequality, and range), aggregations are parsed into functions (sum, avg, min, max, etc.) with their respective columns, ordering into direction (asc, desc) or even aggregate functions, row count limits and so on. Thus, the parsing step converts the LLM inferred operations into a finite set of well-defined operators for the database, i.e., constraint to what is understood by the Tursio engine.

% - parse fragments into meaningful operators \\
% - robust parsers to handle the LLM noise\\
% - more generalized ANTLR logic\\

\vspace{0.2cm}
{\bf Grounding.}
Once parsed, Tursio grounds the operands and literals in the operators into valid ones using the semantic knowledge graph. This involves matching parsed operands and literals to the corresponding tables, columns, aliases, values, and measures in the knowledge graph. Tursio handles proximity matches (e.g., ``cust'' to ``customer'') and semantic matches (e.g., ``revenue'' to ``sales amount'') during grounding to ensure that the operators are valid and meaningful. It uses a combination of hash-based predictions, distance-based search, embedding-based matching, and LLM adjudication to resolve any ambiguities during grounding. We also apply checks for data types, valid aggregates, and other constraints. The grounding step is crucial for ensuring that the generated query plan can be executed correctly in the database.

% - use metadata to ground the parsed operators into valid ones\\
% - handle proximity and semantic matches during grounding\\

\vspace{0.2cm}
{\bf Tree generation.}
Tursio then composes the grounded operators into a tree structure representing the query plan. We follow standard rules from relational algebra to compose operators, ensuring that the tree is well-formed and adheres to the semantics of the database. Tursio handles operator precedence, associativity, and other constraints during tree generation to ensure that the query plan is correct and efficient. The tree structure allows for further transformations and optimizations in the subsequent steps.

\vspace{0.2cm}
{\bf Transformations.}
After generating the initial operator tree from natural language, Tursio applies two critical post-processing stages to ensure query correctness and security. 

{\it Consolidator} component employs a rule-based framework to resolve inconsistencies and redundancies in the operator tree. It iteratively applies transformation rules that qualify based on operator patterns, including: resolving conflicts when columns appear in both GROUP BY and aggregate expressions; aligning aggregate functions between ORDER BY clauses and SELECT aggregates; inferring missing aggregates when GROUP BY operations lack corresponding aggregations; and triggering semantic reasoning fallback modes when interpretation confidence is insufficient. Each rule implements a two-phase approach—first qualifying whether the transformation is applicable, then applying the modification—enabling systematic refinement of the operator structure.

{\it Qualifier} component enforces row-level security (RLS) policies by transforming generated queries to incorporate QUALIFY clauses. It uses pattern matching to identify table references in FROM and JOIN clauses, then wraps qualifying tables with subqueries that embed the filtering expressions. This transformation ensures that security constraints are transparently applied without requiring modifications to the core query generation logic, maintaining separation of concerns between query semantics and access control enforcement.

\vspace{0.2cm}
{\bf SQL generation.}
Finally, we decode the operator tree representations into executable SQL statements. Each operator type implements specialized conversion logic that handles the semantic and syntactic transformations necessary to bridge the gap between the operator abstraction and database-specific SQL syntax. The conversion process also involves reasoning such as temporal interpretation for date-related queries and mapping specialized vocabularies, e.g., mapping ICD codes filters to hierarchical IN clauses. The decoder also performs query optimization by consolidating related conditions (multiple filters on same column maps to AND and different columns to OR), generating SQL clauses that respect the relationships between filtering, grouping, and aggregation operations. Throughout this process, the system ensures referential integrity across query components and conforms to the target database dialect's syntax and conventions.

% \section{Prompt-based Query Rewriting}
% \label{sec:rewriting}

%   - rewriting rules
%   - expressivity

\vspace{0.2cm}
{\bf Prompt-based query rewriting.}
Natural languages give users the power to express themselves and they expect AI to understand them really well. To enhance the expressiveness of generated queries beyond our operator-based generation framework, we employ a post-processing rewrite mechanism. We analyze the initial generated query against the original user intent and available schema information, and attempt to improvise it using advanced SQL constructs that are not easily representable in the operator abstraction. The rewrite mechanism is guided by structured prompts that allow more advanced SQL features including scalar functions, window functions, nested queries, and common table expressions. We provide the contextual information about the database schema and available expressions, and use LLM to generate improved version that maintains structural consistency and security constraints while answering user questions more accurately. We evaluate the rewritten query for both compilation correctness and accuracy, before deciding which version to pick finally.

\vspace{0.2cm}
To summarize, our approach is to successively reduce the ambiguity when planning natural language queries: parse operators using well crafted ANTLR parsers, ground them using the semantic knowledge graph, compose them into well-formed operator trees, and finally rewrite them for better expressivity.

\section{Tursio Search Interface}
\label{sec:interface}

\begin{figure}[!t]
  \includegraphics[width=\columnwidth]{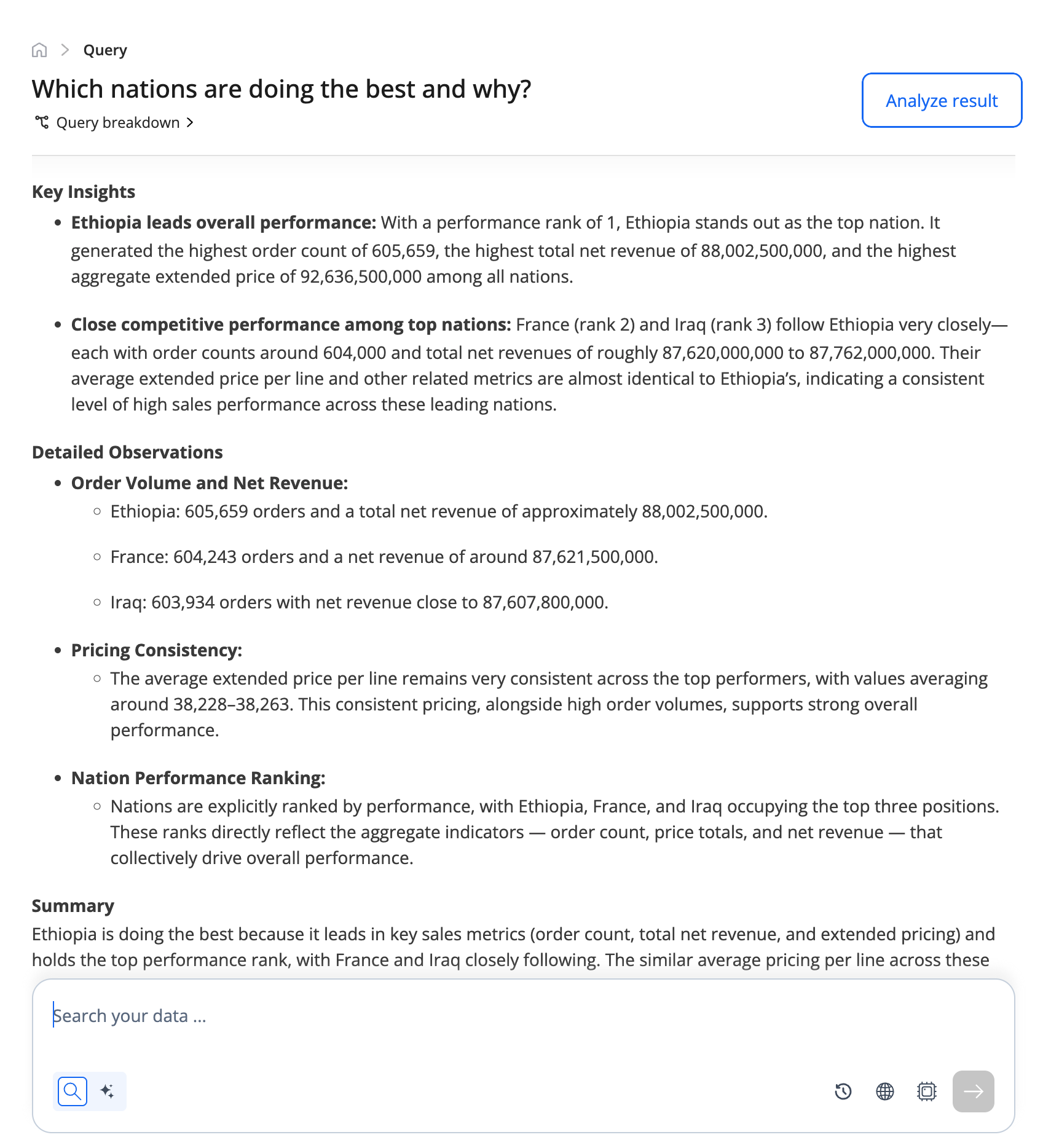}
  \caption{Tursio search interface with a sample question.}
  \Description{Tursio Search Interface}
  \vspace{-0.4cm}
  \label{fig:search-interface}
\end{figure}

Let us now dive deeper into the Tursio search interface that provides a rich set of features for business users to interact with the results. Figure~\ref{fig:search-interface} shows the interface with a sample question ``Which nations are doing the best and why?''. The main features are as follows:

\vspace{0.2cm}
{\bf Query breakdown.} Given that Tursio generates query plans step-by-step, we provide the breakdown of the query plan to users for better explainability. This includes describing the input used, the filters applied, the grouping and aggregations performed, the sorting done, and also which parts of the questions could not be interpreted --- all in user friendly natural language descriptions.

\vspace{0.2cm}
{\bf Auto-prompting.} For expert users, Tursio provides a way to {\it build} their question step-by-step based on the semantic knowledge graph. Users can first pick with the verb (show, list, analyze, etc.), followed by measures, dimensions, and filters (or group by when no value is selected). The scope is automatically updated at every step based on the previous selections from the knowledge graph. Auto-prompted queries are fast and guaranteed to be always correct, i.e., users can construct queries with arbitrarily complex predicates without bothering to have LLMs compile them correctly. 

\vspace{0.2cm}
{\bf Result reasoning.} Tursio reasons the query results to provide responses in natural language. We leverage reasoning models (e.g., o3-mini and o4-mini) to generate concise answers based on the tone of the question, e.g., analyze, summarize, compare, and so on. 
For larger result sets, Tursio creates an agent to build a mini vector index over the results and reason based on that.
We also allow users to provide external context for generating the responses, e.g., latest market trends, political events, and so on. 
Behind the scenes, Tursio pulls the external context using Perplexity APIs and incorporates it for more informed answers.
Finally, users can dig deeper and chat with any given result, e.g., ask more questions about the data, trends, outliers, and so on.

% - hints: similar operations (dimensions, measures, filters, group by, order by, limit), questions

\vspace{0.2cm}
{\bf Tabular result.} Tursio displays the query results in a tabular format for easy exploration. Users can sort columns, resize them, and page through large result sets. We provide options to prune columns, rows, and export the pruned results to Excel for further analysis. To make the data more intuitive, Tursio also provides a statistical summary of the table in natural language.

\vspace{0.2cm}
{\bf Visualization.} Tursio provides a rich set of visualization options, including line, bar, time series, and pie charts. We automatically pick one or more chart types, the dimension and measure columns, chart descriptions, and other settings based on the query result. We further provide users the option to customize the visualizations with different pairs of dimensions and measures.

\vspace{0.2cm}
{\bf Sharing.}
Tursio provides multiple ways to share the results. First of all, users can create a permalink of their result (including data, description, visualizations, etc.) that can be accessed by anyone having the same access control. Users can also export the result into PDF reports for offline sharing. Finally, users can bookmark interesting queries for future references. Bookmarked queries are visible to anyone having access to the same data source.

\vspace{0.2cm}
{\bf History.} Tursio maintains a history of all user queries for easy access. Users can revisit their previous queries, modify them, and re-execute them to get updated results. Furthermore, Tursio recommends interesting questions, based on semantic ranking, from other peoples' history to help users discover new insights.

\vspace{0.2cm}
Overall, Tursio provides a rich set of features in its search interface to make data exploration easy and intuitive for business users.

% (agentic for larger dataset), analyze further
  
%   - hints: similar operations (dimensions, measures, filters, group by, order by, limit), questions
%   - tabular interactions
%   - visualizations
%   - data model
%   - sharing (PDF, link)
%   - saving 
%   - history 
%   - external context
%   - agentic workflow
%   - analysis

\section{Enterprise Readiness}
\label{sec:readiness}

\begin{figure}[!t]
  \includegraphics[width=0.475\textwidth]{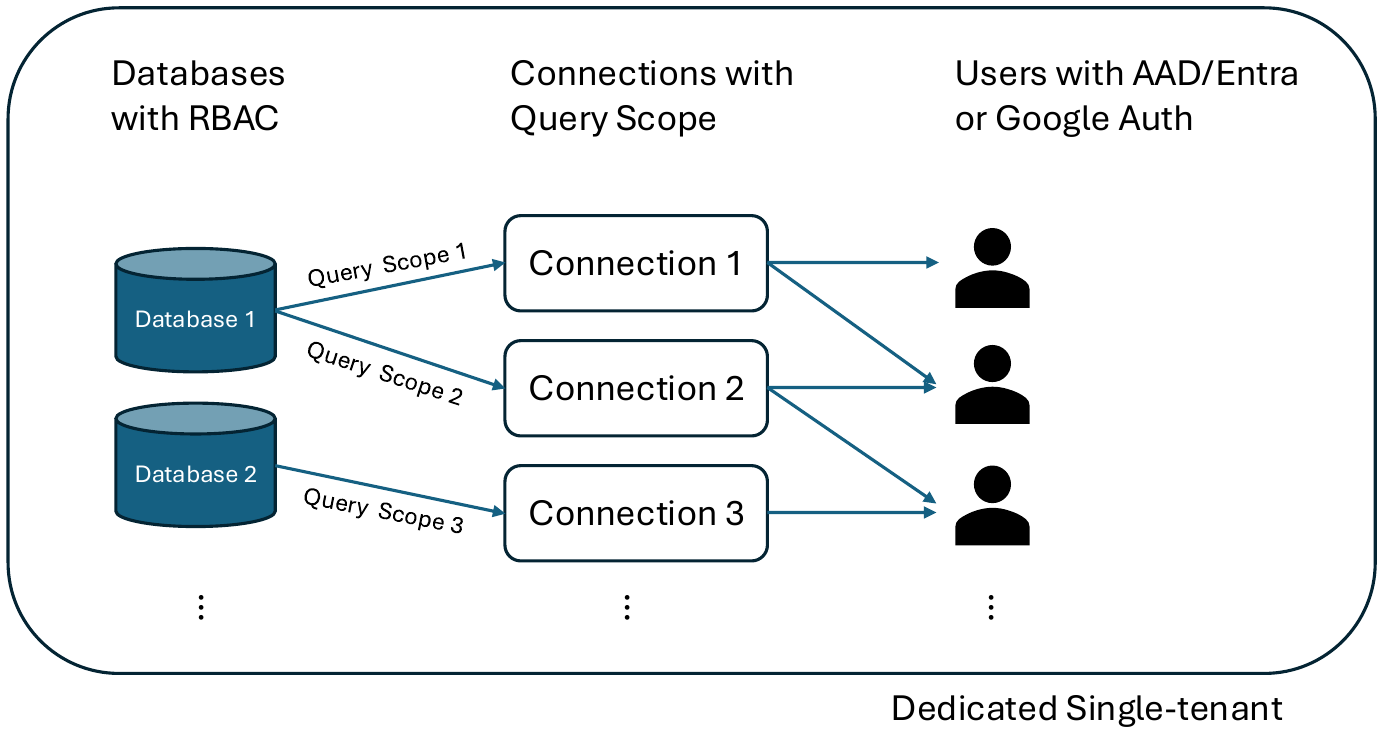}
  \caption{Access control in Tursio.}
  \Description{Access Control}
  \label{fig:access-control}
\end{figure}

Given that Tursio makes a company's sensitive data searchable to a broader audience, enterprise readiness is critical. We discuss this from multiple dimensions below. 

\vspace{0.2cm}
{\bf On-premises.}
First of all, Tursio installs entirely on-premises within a customer's environment, even air-gapped environments. They can run Tursio as standalone Docker container or within their Kubernetes clusters and connect to their databases securely. Tursio supports all major databases and data warehouses and does not require any new infrastructure or data pipelines. It can also be setup as a Azure VM extension (preview) when creating a new VM in the Azure portal. 
On-premises deployment is critical for enterprises in domains with stringent regulatory and compliance requirements, e.g., banking, insurance, healthcare, etc.

% -- data sources \\
% -- deployment (saas, on-prem, embedded)\\

\vspace{0.2cm}
{\bf Security.}
Tursio provides multiple layers of security to protect sensitive data. 
We support configuring HTTPS and SSO for secure access to the Tursio platform.
Users can configure their own OAUTH endpoint for SSO authentication. Tursio also allows creating temporary license keys for secure ad-hoc access.

Figure~\ref{fig:access-control} shows the data access control architecture in Tursio.
Users can connect to multiple data sources, each with its own RBAC. They can also create multiple connections to the same data source with different {\it query scope}, i.e., different sets of tables and columns (column-level security) that can be accessed. Users in Tursio can have Admin, Owner, User, or Viewer roles and they can assigned access to one or more connections.
Finally, Tursio also supports row-level security (RLS) by allowing Admins to provide the filter to inject as CTE for different connections/roles.

\vspace{0.2cm}
{\bf Privacy.}
Privacy is a major concern for enterprises when dealing with AI. Tursio addresses it in multiple ways. First of all, Tursio supports Bring-Your-Own-Key (BYOK) for AI models, e.g., Azure OpenAI, so that enterprises can protect their prompts and apply content filtering as needed.
For enterprises hosting their own models, Tursio provides a stub to validate and plug-in their private LLM endpoints.
To manage costs, Tursio provides a cost estimator for estimating the token costs when scaling the number of search queries.
Apart from model privacy, Tursio also protects data privacy by identifying PII columns and excluding them from training.

\vspace{0.2cm}
{\bf Usability.}
Usability is paramount since that was the main motivation for making databases searchable.
For each role, we allow Admins to configure the query modes (search, auto-prompt), data controls (LIMIT parameter, enable/disable data export), and layout settings (prioritize table, visualizations, SQLs, and disclaimers), or provide custom instructions (formatting, RLS, etc.).
Together with guided tours, our goal is to make Tursio usable by non-expert business users with minimal training.

\vspace{0.2cm}
{\bf Maintenance.}
Users can train the semantic knowledge graph initially and also retrain it periodically as the underlying databases evolve. Tursio detects any manual changes to the knowledge graph and prompts users to retrain it as needed.
To manage the cost of retraining, Tursio supports incremental updates: when schemas change, only the affected tables and their relationships are re-profiled and re-inferred, avoiding the need for full retraining. Tursio tracks schema versions and computes deltas to determine which portions of the knowledge graph require updates, significantly reducing the number of LLM calls for routine schema evolution.
Furthermore, we maintain a migration benchmark to maintain and migrate all prompts within the platform to newer versions of LLMs~\cite{tripathi2025promptmigrationstabilizinggenai}.

\vspace{0.2cm}
{\bf Auditing.}
All interactions with Tursio are logged locally for auditing purposes and accessible to the administrators. They can monitor queries, user satisfaction (thumbs up/down), and can even estimate the return on investment (ROI) based on the time saved with Tursio.
Likewise, all control plane operations are logged and any delete operation is a soft delete, i.e., administrators can always trace back which users ever had access to which data.

\vspace{0.2cm}
Thus, Tursio is designed from the ground up to be enterprise ready, addressing key concerns around deployment, security, privacy, usability, maintenance, and auditing.

\section{Evaluation}
\label{sec:evaluation}

We now present a comprehensive evaluation of our system across production deployments, and both synthetic and realistic academic benchmarks.
We additionally evaluate several auxiliary capabilities critical for enterprise adoption, including column name simplification, personally identifiable information (PII) detection, summary generation, and column description generation.
We describe two case studies of Tursio in production settings, demonstrating its effectiveness in real-world scenarios.
% We conclude with brief observations on ontology alignment and migration support.

\subsection{Datasets}
\label{sec:datasets}
Our evaluation spans four classes of datasets that collectively capture real-world production complexity, research benchmarks, and industry-standard analytical workloads.
This diverse selection ensures that our conclusions are representative of both practical deployments and controlled benchmarking environments.

\vspace{0.2cm}
\textbf{Production Workloads.}
We used internal datasets to evaluate column simplification, PII detection, answer summary generation, and column description tasks.
We evaluate our system on multiple internal datasets originating from live enterprise deployments to verify our model's robustness under non-synthetic, and often noisy data conditions.
These workloads include heterogeneous business domains specifically healthcare, banking, and insurance with varying schema sizes (ranging from tens to hundreds of tables).

\vspace{0.2cm}
\textbf{Synthetic Benchmark: BIRD-DEV.}
BIRD-DEV~\cite{bird_bench} is a widely adopted schema-grounded benchmark designed to evaluate multi-database text-to-SQL systems.
We used all 11 heterogeneous databases, each with varying schema density, key relationships, and domain specificity.
We uploaded all the 11 databases in BIRD-Dev to a Snowflake database instance and trained them independently to build separate semantic knowledge graphs. We evaluated the entire set of $1{,}534$ queries.
We manually annotated the gold table sets for every query and extracted join relationships from explicit schema definitions for our evaluation.

\vspace{0.2cm}
\textbf{Realistic Benchmark: BEAVER.}
We utilized the BEAVER benchmark~\cite{beaver_benchmark}, which represents realistic analytical workloads collected from two large-scale enterprise data warehouses.
Unlike synthetic or academic benchmarks, BEAVER reflects challenges commonly found in operational BI environments such as inconsistent naming conventions, wide tables, and complex join paths.
We evaluated on all query-relevant tables from the two warehouses over $209$ SQL queries. For BEAVER, we included only the query-relevant tables across both warehouses to match the benchmark's intended evaluation scope.
To prepare the dataset for evaluating joins, we relied on the declared primary-key/foreign-key join relationships in the CSAIL datasets.
For the DW dataset, where join relationship metadata is not available, we extracted join paths using all unique join relationships that were explicitly referenced in the benchmark queries.

\vspace{0.2cm}
\textbf{Industry Standard Benchmark: TPC-H.}
To compare our system on established decision support systems, we also included the TPC-H benchmark~\cite{tpch_benchmark}.
We used all the eight core tables of TPC-H to assess our model's robustness and scalability in TPC-H like decision-support workloads.

\begin{figure}[t]
    \centering
    \includegraphics[width=\columnwidth]{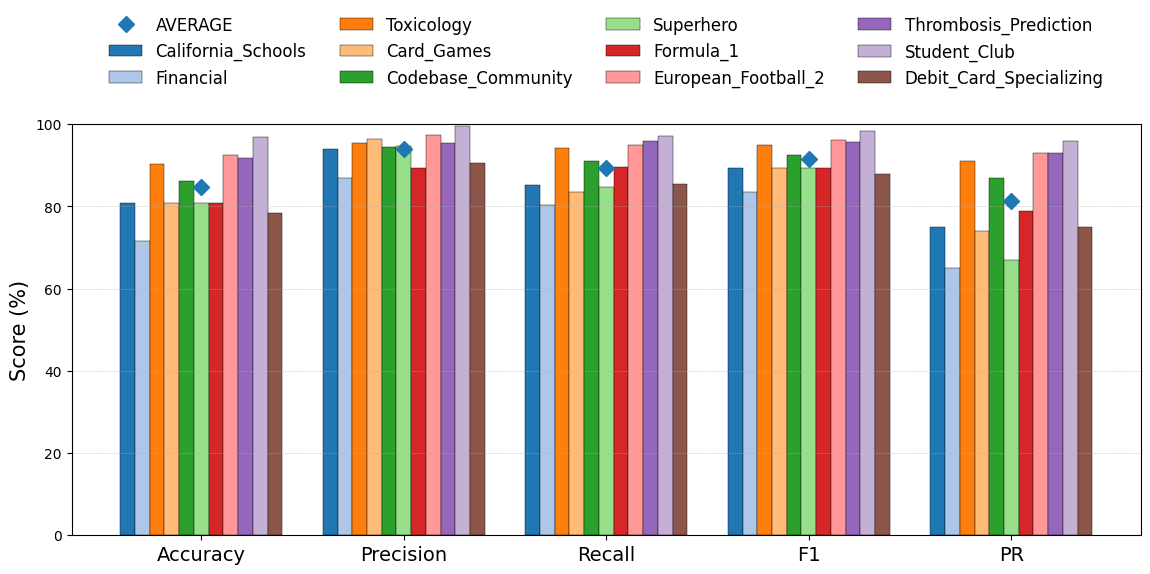}
    \caption{\centering Table Retrieval Performance (BIRD)}
    \Description{Table Retrieval Performance (BIRD)}
    \label{fig:table_retrieval_bird}
\end{figure}

\subsection{Table Retrieval}

\begin{figure}[t!]
    \centering
    \includegraphics[width=\columnwidth]{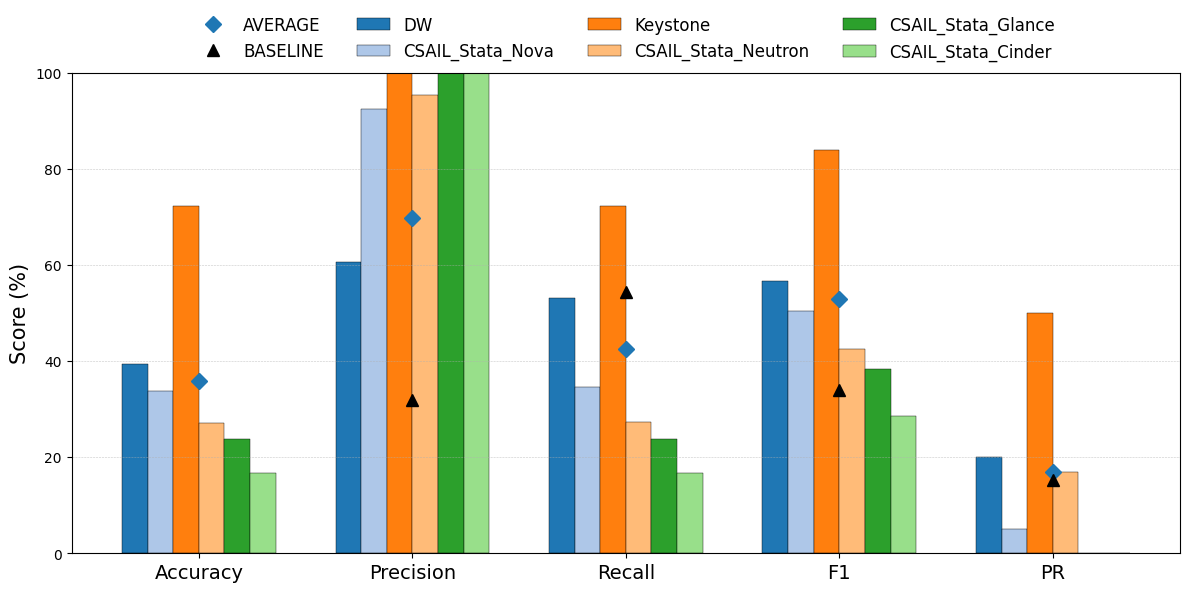}
    \caption{Table Retrieval Performance (BEAVER)}
    \Description{Table Retrieval Performance (BEAVER)}
    \label{fig:table_retrieval_beaver}
\end{figure}

First of all, we evaluate table retrieval, which measures whether the system correctly identifies all tables required to answer a query. Following the same approach as in prior work~\cite{beaver_benchmark}, we evaluate:
\begin{itemize}
    \item Precision — fraction of retrieved tables that are correct,
    \item Recall — fraction of relevant tables successfully retrieved,
    \item F1 score, and
    \item Perfect-Recall — proportion of queries for which all gold tables are retrieved.
\end{itemize}

Perfect-Recall is particularly important in production scenarios, where missing even a single table often leads to syntactically valid but semantically incorrect SQL.
Figures~\ref{fig:table_retrieval_bird} and~\ref{fig:table_retrieval_beaver} show the table retrieval results for BIRD-DEV and BEAVER.

The results for BIRD-DEV show that Tursio achieves high overall accuracy ($85\%$) with balanced precision ($94\%$), recall ($89\%$), F1 ($92\%$), and Perfect-Recall ($81\%$).
For the BEAVER dataset family, we see a lower recall but reasonable precision, suggesting that while the model avoids retrieving irrelevant tables, additional signal (e.g., extended metadata) could further improve coverage.
Still, compared to the best-performing BEAVER baseline (Precision $32\%$, Recall $54.4\%$, F1 $33.9\%$, Perfect-Recall $15\%$), Tursio delivers substantial gains: in Precision and F1 scores ($91\%$ and $50\%$ respectively), marginal improvement in Perfect-Recall to $15.33\%$. The lower recall of $38\%$ indicates that our model is more conservative in retrieving tables, which may be a trade-off to maintain high precision in the face of noisy or ambiguous schemas.
Here we compare against best-performing baselines for BEAVER which include the best of all the models evaluated in the BEAVER paper, across both the top-10 and top-5 retrieval performance. In our table retrieval algorithm, we retrieve the exact tables that are required to answer the query, which is a more stringent requirement than retrieving a superset of relevant tables. This leads to higher precision but can reduce recall if the model fails to identify all necessary tables.
Overall, the results demonstrate that Tursio achieves strong table retrieval performance across both the benchmark datasets.

\subsection{Join Inference}

\begin{figure}[t!]
    \centering
    \includegraphics[width=\columnwidth]{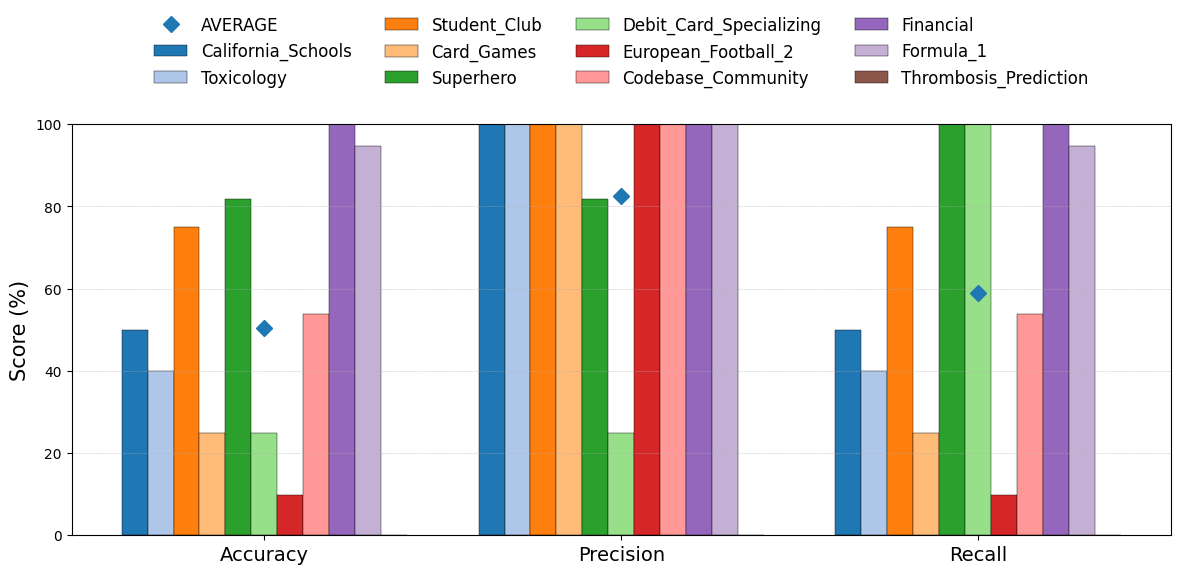}
    \caption{Join Inference Performance (BIRD)}
    \Description{Join Inference Performance (BIRD)}
    \label{fig:join-inference-bird}
\end{figure}

We now evaluate Tursio's ability to discover relationships between tables during the semantic model construction, i.e., we measure the accuracy in inferring primary--foreign key relationships.
We report \emph{precision}, \emph{recall}, and \emph{F1 score}, which capture:
\begin{itemize}
    \item The proportion of inferred joins that are correct (precision),
    \item The proportion of ground-truth joins recovered (recall), and
    \item The overall quality of join inference (F1).
\end{itemize}

\begin{figure}[t!]
    \centering
    \includegraphics[width=\columnwidth]{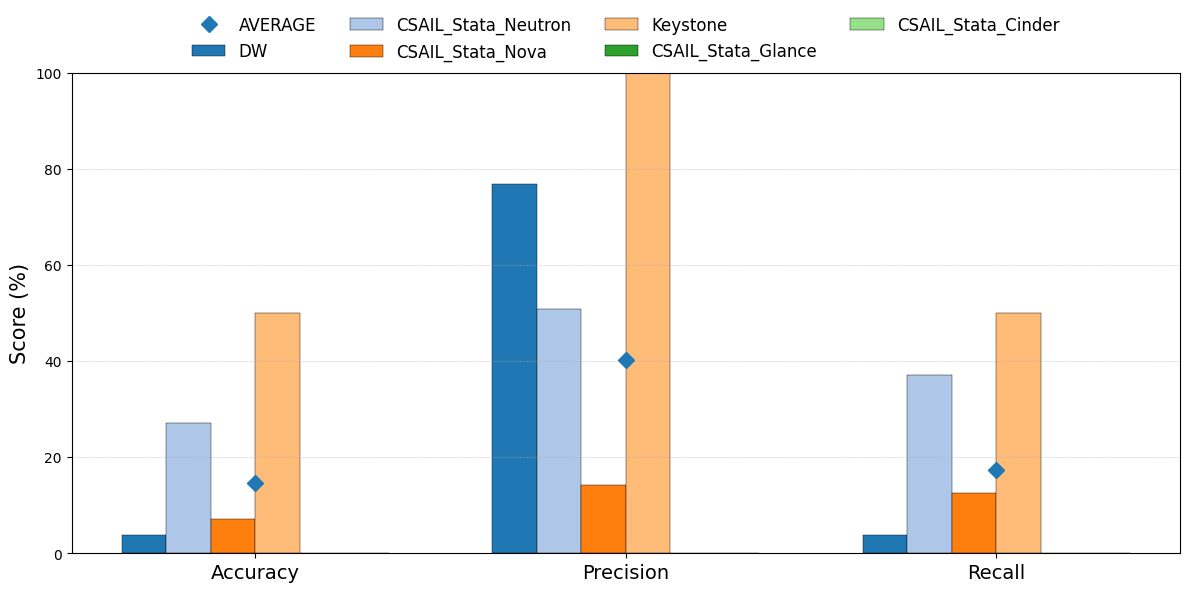}
    \caption{Join Inference Performance (BEAVER)}
    \Description{Join Inference Performance (BEAVER)}
    \label{fig:join-inference-beaver}
\end{figure}

\begin{figure}[t!]
    \centering
    \includegraphics[width=\columnwidth]{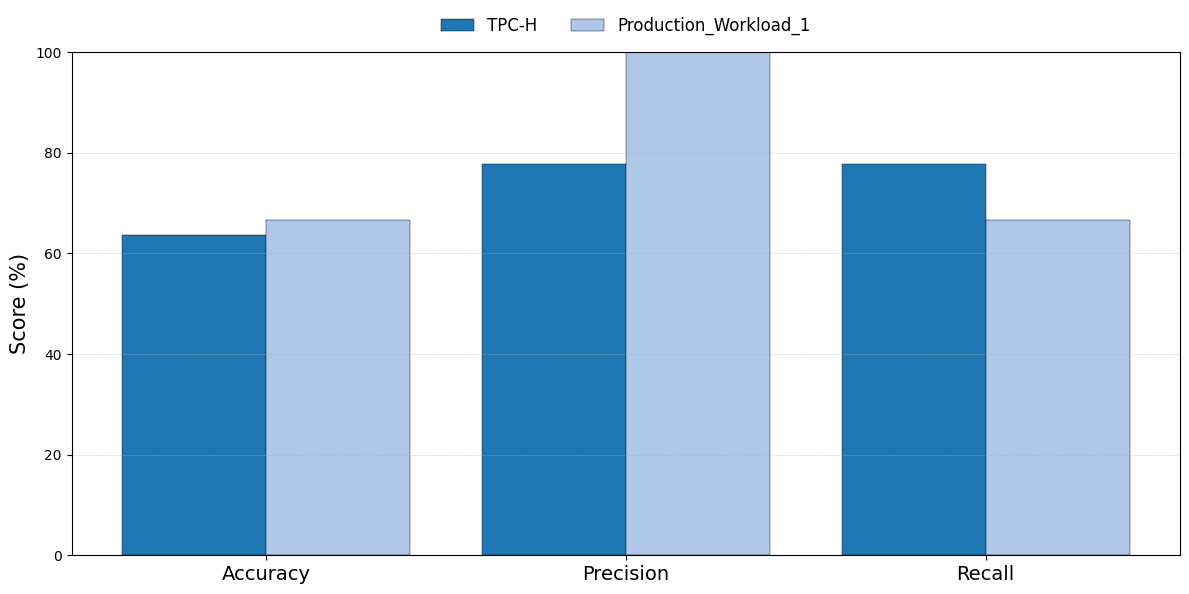}
    \caption{Join Inference Performance (Other Datasets)}
    \Description{Join Inference Performance (Other Datasets)}
    \label{fig:join-inference-tpch_prod_workload1}
\end{figure}

\begin{figure*}[t]
    \centering
    \includegraphics[width=\textwidth]{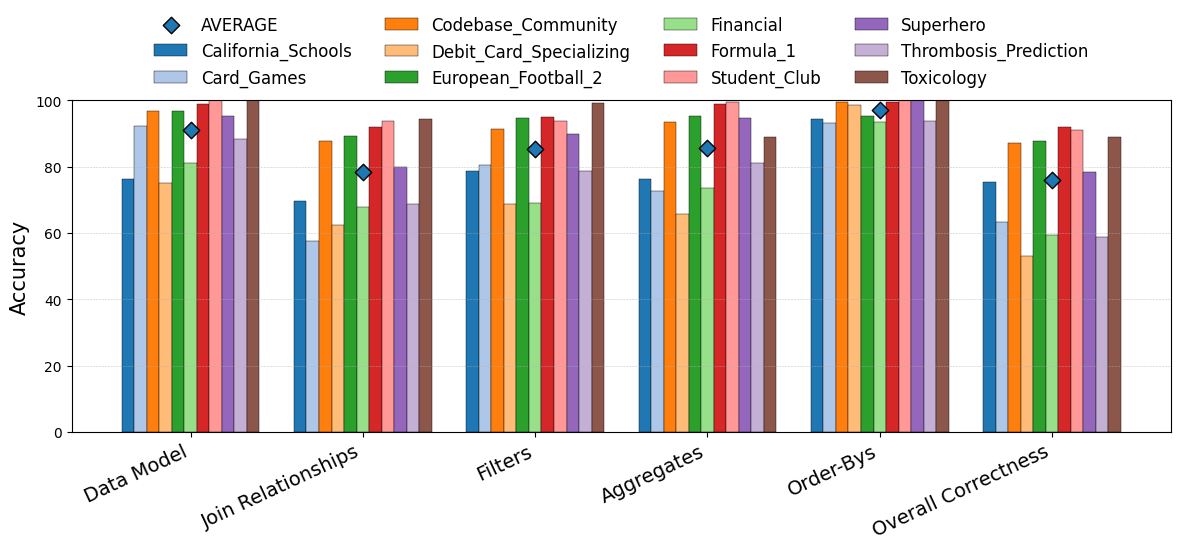}
    \caption{\centering SQL Structural Accuracy (BIRD)}
    \Description{SQL Structural Accuracy (BIRD)}
    \label{fig:sql_sttructural_acc_bird}
\end{figure*}

For BIRD-DEV and TPC-H, the ground truth consists of the full set of joins defined in the official schema metadata, including both explicit primary--foreign key pairs and additional schema-declared referential constraints.
For the BEAVER benchmark, we use all annotated join pairs provided for both warehouses. For the internal production workload, join annotations were curated by domain experts.
During evaluation, we compare these ground-truth pairs with the join connections inferred by our semantic knowledge-graph construction. We report the resulting precision, recall, and F1 scores in Figures~\ref{fig:join-inference-bird}, ~\ref{fig:join-inference-beaver}, and~\ref{fig:join-inference-tpch_prod_workload1}.
The BEAVER schemas represent the most challenging setting, exhibiting limited primary--foreign key discipline. Consequently, our model achieves low average Accuracy ($14.7\%$), Precision ($40\%$), and Recall ($17.3\%$), as it struggles to distinguish true join paths from semantically plausible alternatives in loosely structured schemas.
In contrast, performance on public benchmark datasets is substantially stronger. On BIRD, the model attains an average Accuracy of $50.46\%$, Precision of $82.44\%$, and Recall of $58.93\%$. These improvements stem from the clearer relational structure and more consistent key constraints, which enable recovery of most valid join relationships with high precision. Nonetheless, imperfect normalization in some cases prevents us from having a better recall.

TPC-H and production workloads further illustrate the model's upper-bound performance in well-engineered environments. These schemas are normalized and stable, allowing the model to achieve its strongest results—Accuracy exceeding $63\%$, Precision reaching $100\%$, and Recall between $67\%$ and $78\%$.

\vspace{0.2cm}
Overall, the above results demonstrate that join-inference performance varies substantially across dataset families, reflecting the model's sensitivity to schema regularity and design discipline. Schemas that are well-structured yield consistently high-quality inference, whereas irregular or weakly constrained schemas pose significant challenges.

\subsection{Query Generation}

We now evaluate the SQL generation in Tursio.
Since Tursio does not translate questions to SQLs but rather generates descriptive answers, we focus on evaluating the structural correctness using an LLM-as-an-evaluator protocol that compares predicted SQL queries against the annotated ground-truth queries.
The evaluator examines the predicted SQL and produces a structured assessment across the following dimensions:

\begin{itemize}
    \item {Data Model Alignment:}  Assesses whether the predicted query references the correct tables and columns as defined in the database schema. Basically, this checks whether the query operates on semantically relevant entities, avoids hallucinated attributes, and follows the referential structure implied by the schema.
    \item {Join Logic:} Evaluates whether the predicted join structure matches the ground truth in terms of both join partners and join predicates. The evaluator verifies correctness of join types, key usage, and multi-hop join paths, penalizing missing joins, spurious joins, or incorrect join conditions.
    \item {Filters:} Measures fidelity of selection predicates. The evaluator verifies that the predicted WHERE conditions capture the correct attributes, operators, literal values, and logical combinations (e.g., conjunctions or disjunctions), and that the semantics of the filters match the ground truth without over- or under-specification.
     \item {Measures and Aggregates:} Checks whether aggregate functions (e.g., \texttt{SUM}, \texttt{COUNT}, \texttt{AVG}), grouping keys, and computed measures are correctly reproduced. This evaluation ensures that the grouping granularity matches the reference query and that aggregated outputs reflect the intended analytical context.
     \item {Order-By Conditions:} Assesses correctness of ordering specifications, including ordering attributes, ascending/descending direction, composite orderings, and alignment with the ground-truth intent. The evaluator distinguishes between necessary ordering for correctness and incidental or stylistic ordering differences.
\end{itemize}

In addition to per-component judgments, the evaluator outputs a categorical \textit{Overall Match} score with one of three values: Correct, Partial, or Incorrect.
These labels reflect strict adherence to the ground-truth semantics; the evaluation prompt explicitly instructs the model to avoid false positives and false negatives.

Figure~\ref{fig:sql_sttructural_acc_bird} shows the component-wise structural evaluation on the BIRD-DEV benchmark. We can see a strong alignment between the predicted and the reference SQLs across most dimensions:
\begin{itemize}
    \item \textbf{Data Model Alignment:} We achieve an accuracy of $90\%$, indicating strong consistency between the predicted schema mappings and the ground-truth database structure.
    \item \textbf{Ordering:} ORDER BY clauses are identified with near-perfect accuracy ($97\%$), underscoring a robust capacity to correctly infer the ranking indicated by users.
    \item \textbf{Filters:} Filter predicates exceed $85\%$ accuracy, reflecting a robust extraction of analytic and logical requirements from natural language input.
    \item \textbf{Aggregation:} Aggregate clauses also exceed $85\%$ accuracy, showing reliable capture of analytic intent and grouping semantics.
    \item \textbf{Join Logic:} Join structure accuracy ($78\%$) suggests errors are concentrated in multi-table, multi-hop join scenarios and in queries requiring complex compositional reasoning (e.g., nested subqueries with correlated filters).
    \item \textbf{Full-query Equivalence:} Overall we have an accuracy of $76\%$ for full SQL structural equivalence.
\end{itemize}
Despite these remaining challenges, the LLM-as-evaluator protocol confirms that predicted SQL retains high structural fidelity across most core aspects of SQL generation, evidenced by consistently strong performance over varied database domains.

\subsection{Column Name Simplification}

\begin{figure}[t!]
    \centering
    \includegraphics[width=\columnwidth]{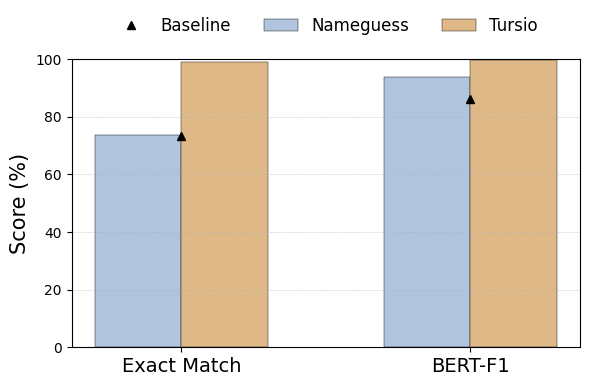}
    \caption{Column Name Simplification}
    \vspace{-0.4cm}
    \Description{Column Name Simplification}
    \label{fig:column_name_simplification}
\end{figure}

We now evaluate column name simplification using both the NameGuess benchmark~\cite{name_guess} and an internal production workload.
For NameGuess, we report exact-match and BERT-F1 scores, which are standard metrics for assessing the semantic and lexical alignment of expanded column names. Figure~\ref{fig:column_name_simplification} shows the results.
On the NameGuess dataset, our approach achieves $73.8\%$ exact match and 93.88\% BERT-F1, indicating improved semantic alignment over the baseline, though exact lexical matches remain challenging due to the synthetic nature of the benchmark's abbreviations~\cite{name_guess}.

In contrast, evaluation on the internal Tursio production workload yields substantially higher scores, with $99.0\%$ exact match and $99.69\%$ BERT-F1.
These near-perfect results suggest that, in real-world datasets, our method captures a robust understanding of contextual and semantic relationships, outperforming benchmarks that rely on synthetic abbreviations.
Our human evaluation further confirms that we achieve high semantic adequacy and user acceptability, with simplified names rated for clarity, interpretability, and domain appropriateness.
The results, summarized in Figure~\ref{fig:column_name_simplification}, highlight clear improvements in performance for both benchmark and production settings.

\subsection{Auxiliary Capabilities}

Finally, we evaluate the auxiliary tasks that support system features required in production use cases.
Table~\ref{tab:aux-tasks} summarizes the results for auxiliary tasks.

% \begin{itemize}
\vspace{0.2cm}
{\bf Summary Quality.}
To assess the quality of automatically generated summaries, we employ an LLM-as-a-grader evaluation protocol using GPT-4.1 and GPT-5.
The grader is provided with a structured rubric that categorizes each summary into one of four quality levels: Excellent, Good, Average, or Poor.
The rubric evaluates five dimensions—clarity and coherence, conciseness, structure, factuality, and length appropriateness—and requires the model to output both a categorical rating and a short justification.
Each categorical label is mapped to a numerical score using a fixed scale (Excellent = 1, Good = 2, Average = 3, Poor = 4).
For every summary, both models independently produce a score, and we report the average of their individual scores.
The summary quality evaluation, via dual LLM grading, achieves an average score of 2.2, suggesting high-quality structured summaries.

\vspace{0.2cm}
{\bf PII Detection.} We compute the precision and recall over labeled PII entities to evaluate whether the system successfully detects sensitive columns while minimizing false positives.
Our results show that PII detection achieves strong recall ($95\%$) and precision ($91\%$), which is suitable for production-grade privacy constraints.

\vspace{0.2cm}
{\bf Column Descriptions.} Our internal evaluators assign binary (0/1) relevance scores reflecting whether the generated column descriptions accurately and succinctly capture the semantic meaning.
Our human relevance ratings for column descriptions ($4.4/5$) demonstrate strong interpretability and user-facing clarity.

% \end{itemize}

\begin{table}[t!]
\centering
\small
\begin{tabular}{lcc}
\toprule
\textbf{Task} & \textbf{Metric} & \textbf{Score} \\
\midrule
PII Detection                    & Recall      & 95\% \\
                                 & Precision   & 91\% \\
Summary Quality (LLM-Graded)     & Average Score & 2.2 \\
Column Descriptions              & Human Relevance Score & 4.4 / 5 \\
\bottomrule
\end{tabular}
\vspace{0.2cm}
\caption{Performance metrics for auxiliary semantic tasks.}
\vspace{-0.3cm}
\label{tab:aux-tasks}
\end{table}

\vspace{0.2cm}
{\bf Domain Ontology.}
Tursio has support for domain ontology as part of query processing.
% In ``Searching Clinical Data Using Generative AI''~\cite{clinical_search_tursio}, we present SearchAI, a generative AI approach for
For example, we allow searching clinical data that leverages hierarchical models to respect coding ontologies, e.g., ICD-10~\cite{clinical_search_tursio}.
The system decomposes natural language queries into Boolean logic, traverses the code hierarchy predictively, and ensures all relevant paths are reachable.
% The paper details the challenges of automating clinical code lookup, including
This is challenging due to the complexity of large hierarchies, ambiguous descriptions, and inconsistent parent-child relationships.
% Through a series of experiments, we show that SearchAI
Our algorithms clearly outperform default hierarchical traversals in accuracy, robustness, performance, and scalability, making clinical data more accessible and streamlining the workflows for healthcare professionals.

\vspace{0.2cm}
{\bf Prompt Migration.}
The rapid evolution of Large Language Models (LLMs) poses significant challenges for the stability and reliability of GenAI applications. 
% In ``Stabilizing GenAI Applications with Evolving Large Language Models'',
Therefore, we introduce the concept of \textit{prompt migration} as a systematic approach to maintain application consistency amid changing LLMs~\cite{tripathi2025promptmigrationstabilizinggenai}.
% Using the Tursio enterprise search application as a case study, we analyze the impact of
This would help migrate Tursio over successive GPT model versions (GPT 4.1, 5.0, 5.1, and so on).
We also came up with a migration framework—including prompt redesign and a migration testbed—and demonstrate how these techniques restore application reliability lost due to model drift.
This also highlights the importance of prompt lifecycle management and robust testing for AI-powered systems.

% to ensure dependable GenAI-powered business applications, providing practical lessons for developers and organizations navigating the dynamic landscape of LLM evolution.

\vspace{0.2cm}
Across multiple auxiliary tasks, Tursio is designed for robust performance, high fidelity, and user acceptability. Overall, this reflects the suitability of Tursio for end-to-end analytical workflows in enterprise environments.

\subsection{Case Studies}

We now present two case studies of Tursio in action on real-world production workloads.

\vspace{0.2cm}
{\bf Case study 1: Serving customer tickets.}
A large home warranty company wants to improve their customer experience with better ticket handling.
Their support platform generates templated queries that must fetch relevant data, analyze the typical resolution, apply business rules to ensure the promised customer agreements are respected, and finally generate a response to resolve the request.
The underlying schema spans over several tables covering warranty contracts, service history, customer profiles, and coverage terms, with domain-specific terminology (e.g., ``covered items'', ``courtesy amount'') that general-purpose LLMs would not understand out of the box.
Tursio's semantic knowledge graph automatically mapped these domain terms to the correct columns and inferred the join paths between contract, service, and customer tables. Custom measures were defined to encode business rules such as coverage eligibility and service fee calculations, ensuring that the generated responses respect the contractual agreements.
For each response generated by Tursio, a human agent verifies its correctness (marking it as helpful or not helpful) before sending it to the customer.
Table~\ref{tab:case-study-1} shows the human-evaluated accuracy over a recent six-week period of production usage. Serving hundreds of queries per week, Tursio achieves over $96\%$ raw accuracy and $100\%$ effective accuracy (excluding cases when insufficient information was provided as input) consistently.
The gap between raw and effective accuracy is attributed entirely to incomplete input tickets that lack required details (e.g., missing diagnostic details), rather than errors in query planning or result generation.

\vspace{0.2cm}
{\bf Case study 2: Dynamic dashboard querying.}
A supply chain and planning company wants to allow its users to dynamically query their data along various planning points.
Their existing dashboards are static and do not allow users to ask ad-hoc questions, and they require natural language queries to be interpreted based on their underlying data model.
The data model contains wide planning tables (50+ columns) with complex hierarchical dimensions (product categories, geographic regions, time periods) and domain-specific metrics.
Tursio's join inference was critical for connecting the planning dimension tables, while auto-generated sample questions provided grounding for the diverse ways domain experts phrase analytical queries over these dimensions.
They ran an exhaustive validation using a set of 192 queries crafted by their domain experts over their data model.
Table~\ref{tab:case-study-2} shows the accuracy of Tursio in terms of compilation and validation over these queries.
Tursio achieves $86\%$ compilation accuracy, which goes up to $88\%$ when excluding cases with missing data.
Furthermore, when validating queries while excluding those with missing data or metrics, Tursio achieves $94\%$ validation accuracy.
An analysis of the failure scenarios revealed that the majority involved metrics not yet modeled in the knowledge graph or queries requiring multi-step reasoning across temporal hierarchies---areas targeted for improvement in future iterations.

\begin{table}[t]
\centering
\small
\begin{tabular}{lrrrrrr}
\toprule
\textbf{Metric} & \textbf{W1} & \textbf{W2} & \textbf{W3} & \textbf{W4} & \textbf{W5} & \textbf{W6}\\
\midrule
Queries & $681$ & $667$ & $641$ & $662$ & $764$ & $739$ \\
Accuracy (raw) & $96.3\%$ & $97.3\%$ & $97.9\%$ & $96.2\%$ & $97.4\%$ & $97.1\%$ \\
Accuracy (eff) & $100\%$ & $100\%$ & $100\%$ & $100\%$ & $100\%$ & $100\%$ \\
\bottomrule
\end{tabular}
\vspace{0.2cm}
\caption{Case study 1: Serving customer tickets.}
\label{tab:case-study-1}
\end{table}

\begin{table}[t]
\centering
\small
\begin{tabular}{lr}
\toprule
\textbf{Metric} & \textbf{Accuracy}\\
\midrule
Compilation & $86\%$ \\
Compilation (excl. missing data) & $88\%$ \\
Validation (excl. missing data, metrics) & $94\%$ \\
\bottomrule
\end{tabular}
\vspace{0.2cm}
\caption{Case study 2: Dynamic dashboard querying.}
\vspace{-0.4cm}
\label{tab:case-study-2}
\end{table}

% \subsection{Conclusion}
% \label{sec:conclusion}

\vspace{0.2cm}
To summarize, across all evaluation tasks, Tursio consistently demonstrates strong semantic understanding, robust structural reasoning, and high-quality auxiliary capabilities.
Collectively, these results show that our system delivers high structural accuracy in SQL generation, reliable semantic interpretations, and robust auxiliary capabilities, particularly on datasets with well-defined schemas.
The combination of strong table retrieval, precise join reasoning, and high-quality summarization establishes the system as a practical and effective solution for both production workloads and research benchmarks.

\section{Related Work \& Road Ahead}
\label{sec:roadahead}

\vspace{0.2cm}
{\bf Text-to-SQL.}
Translating natural language to SQL has been extensively studied, with benchmarks like Spider~\cite{spider} and BIRD~\cite{bird_bench} driving rapid progress. Recent LLM-based approaches include DIN-SQL~\cite{dinsql}, which decomposes the task into sub-problems, DAIL-SQL~\cite{dailsql}, which optimizes example selection for in-context learning, CHESS~\cite{chess}, which adds contextual retrieval and schema pruning, and MAC-SQL~\cite{macsql}, which employs a multi-agent architecture. While these approaches focus on generating SQL from natural language, they assume a given schema and do not address the upstream challenge of automatically inferring semantic context from the raw database. Tursio targets the broader end-to-end problem: from automatic semantic modeling to query planning to delivering natural language answers rather than SQL. Furthermore, Tursio's multi-step planning with deterministic grounding constrains LLM outputs at each stage, unlike end-to-end approaches that rely solely on the model to produce valid SQL in a single generation step.

\vspace{0.2cm}
{\bf Data Discovery.}
Discovering relationships across tables is central to data integration and search. Aurum~\cite{aurum} builds an enterprise knowledge graph for data discovery using profiling signals. Valentine~\cite{valentine} provides a comprehensive benchmark for schema matching techniques. Starmie~\cite{starmie} uses pre-trained language models for column-level data discovery in data lakes. Tursio's join inference builds on inclusion dependency discovery~\cite{Rostin2009AML} and extends it with LLM-based validation, combining statistical signals (approximate count distinct overlap) and semantic reasoning (LLM as judge) for robust primary-key/foreign-key detection. This hybrid approach allows Tursio to scale join inference to larger schemas while maintaining high precision.

\vspace{0.2cm}
{\bf Semantic Layers and NL Interfaces.}
Semantic layers such as dbt Semantic Layer, LookML, and Cube provide business-friendly abstractions over databases but require manual modeling and ongoing maintenance. Tursio automates this step using LLM-powered inference over the raw database schema and data profiles. Earlier NL database interfaces like NaLIR~\cite{nalir} used interactive query tree disambiguation over dependency parses, while neural semantic parsing approaches like RAT-SQL~\cite{ratsql} pioneered relation-aware schema encoding and linking. Tursio combines ideas from both traditions: it uses schema-aware encoding via the semantic knowledge graph (similar to neural approaches) while preserving the ability to decompose and validate query plans step-by-step (similar to rule-based approaches), resulting in a system that is both accurate and explainable.

% \section{Accuracy \& Reliability}

% -- accuracy numbers \\
% -- explainability\\
% -- interpreted / non-interpreted\\
% -- hints: similar operations (dimensions, measures, filters, group by, order by, limit)\\
% -- hints: similar questions \\
% -- no answers for unclear questions\\

% \section{Automated Semantic Modeling}

% - names, \\
% - types (data, value), \\
% - aspects (entity, currency, precision), \\
% - joins, \\
% - aggregators, \\
% - ontology (ICD, timeseries, etc.), \\
% - aliases \\

% \section{Prompt Query language}

% Prompting is a key skill in generative ai applications and the above query processing good for people who know or are willing to learn how to prompt;\\
% Unfortunately, most people want to start typing off the hook and end up with accuracies anywhere from 85-95\%.\\ 
% Unlike text-to-SQL, they still get consistent answers with explainability, and even no answers when the question is not clear.\\

% \noindent -- automatic semantic modeling\\
% -- auto join algorithm (with results)\\
% -- name expansion (with results)\\
% -- hierarchical ontologies (ICD, drugs, etc.)\\
% -- data types, value types, aspects, etc.\\
% -- automatic context generation\\
% -- join graph traversal\\

% \section{Result Processing}

% -- reasoning models\\
% -- follow-up reasoning\\
% -- streaming latency results \\
% -- agentic reasoning (with size and latency results)\\
% -- tabular interactions\\
% -- visualizations \\
% -- data models\\
% -- sharing (PDF, link, prompt, history)\\
% -- external context\\
% -- agentic workflow\\

% \section{The Road Ahead}
% \label{sec:roadahead}

\vspace{0.2cm}
Tursio has been under development over the last three years and has been serving customer workloads for over a year now. While we have made significant progress, there are still many challenges and opportunities ahead. Below, we discuss some of the key areas for future work.

\vspace{0.2cm}
{\bf Newer world.} Language models have come a long way, from sparking curiosity to creating wow moments to slowly becoming a part of everyday life. They have transformed the way we interact with machines, making AI a technology for the masses. While this is exciting, it has also raised the bar for accuracy, reliability, and usability. Future data systems will inevitably reflect this newer reality and Tursio is a part of that process.

\vspace{0.2cm}
{\bf Noise cancellation.}
General purpose LLMs are trained on public data and getting them to focus on private enterprise data is a challenge. Tursio approach is to successively reduce the ambiguity at every step, from parsing to grounding to tree generation to rewriting, so that the final response is as accurate as possible. Future work will focus on further reducing the noise with self-tuning mechanisms using the query history.

\vspace{0.2cm}
{\bf Search for all data.}
Modern users don't care where the data lives, they just want to search it and get answers.
While Tursio currently supports both SQL and NoSQL data stores, our evaluation has primarily focused on relational databases. Extending the evaluation to NoSQL systems (e.g., Cassandra, CosmosDB) and systematically assessing how query planning generalizes across database dialects is an important next step.
Beyond existing connectors, future work will focus on connecting to unstructured and even web data. Identifying possible ways to link structured and unstructured data together will be a key research challenge.

\vspace{0.2cm}
{\bf Feeding agents.}
Agents are becoming increasingly popular for automating business workflows. However, they need a reliable source of truth to get accurate facts. Tursio can serve as that source of truth layer, providing consistently accurate facts for both humans and agents alike. Future work will focus on integrating Tursio with popular agent frameworks and enabling seamless interactions between them.

% -- Reduce ambiguity\\
% -- AI is on an express pace of development\\  
% -- at the same time, it must work in practice\\
% -- brand new query engine to get facts from structured data\\
% -- Tursio has been serving customer workloads over a year now\\
% -- direct consumption: obvious use case is for people who were unnecessarily relying on dashboards and reports earlier\\
% -- but can also be embedded into existing business workflows or newly emerging agentic workflows\\
% -- the source of truth layer providing consistently accurate facts for human or agent alike\\

\section{Conclusion}

Large language models (LLMs) are democratizing domains that were once the exclusive domain of experts. Data is one such domain where business users have long struggled to access and analyze data without relying on experts.
In this paper, we presented Tursio, a generative AI powered data search engine that makes databases searchable using natural language. Tursio leverages LLMs to infer a semantic knowledge graph over the underlying databases, contextualizes user intent to relevant portions of the knowledge graph, generates query plans in multiple steps, and provides a rich search interface for business users to interact with the results. We discussed the enterprise readiness of Tursio from multiple dimensions, including deployment, security, privacy, usability, maintenance, and auditing. Finally, we presented a detailed evaluation of Tursio on multiple benchmarks and real-world customer workloads. Our results show that Tursio achieves high accuracy and reliability while being easy to use and enterprise ready. We believe that Tursio represents a significant step towards making data more accessible and usable for business users.

\begin{acks}
We thank our colleagues at Tursio for their valuable feedback and support throughout the development of this work. We also sincerely acknowledge our design partners who helped shape the product.
\end{acks}

\balance
\bibliographystyle{ACM-Reference-Format}
\bibliography{references}

% \appendix









\end{document}